\documentclass{article}
\usepackage{graphicx} 
\graphicspath{ {./images/} }
\usepackage{caption}
\usepackage{subcaption}

\usepackage{amsmath}
\usepackage{multirow}
\usepackage{authblk}
\usepackage[letterpaper,top=2cm,bottom=2cm,left=3cm,right=3cm,marginparwidth=1.75cm]{geometry}

\usepackage[colorlinks=true, allcolors=blue]{hyperref}
\usepackage[natbibapa]{apacite}

\title{Remaining Useful Life Modelling with an Escalator Health Condition Analytic System}
\author[1,2]{Inez M. ZWETSLOOT}
\author[1]{Yu LIN}
\author[1]{Jiaqi QIU}
\author[2]{Lishuai LI}
\author[3]{LEE William Ka Fai}
\author[3]{YEUNG Edmond Yin San}
\author[3]{YEUNG Colman Yiu Wah}
\author[3]{WONG Chris Chun Long }

\affil[1]{Systems Engineering, City University of Hong Kong}
\affil[2]{School of Data Science, City University of Hong Kong}
\affil[3]{MTR Corporation Limited}

\begin{document}

\maketitle

\begin{abstract}
An escalator's refurbishment is usually linked with its design life as recommended by the manufacturer. However, the actual useful life of an escalator should be determined by its operating condition which is affected by the runtime, workload, maintenance quality, vibration, etc., rather than age only. The objective of this project is to develop a comprehensive health condition analytic system for escalators to support refurbishment decisions. The analytic system consists of four parts: 1) online data gathering and processing; 2) a dashboard for condition monitoring; 3) a health index model; and 4) remaining useful life prediction. The results can be used for a) predicting the remaining useful life of the escalators, in order to support asset replacement planning and b) monitoring the real-time condition of escalators; including alerts when vibration exceeds the threshold and signal diagnosis, giving an indication of possible root cause (components) of the alert signal.
\end{abstract}

\section{Introduction}

An escalator is a moving staircase that carries people between floors of a building. Escalators are an important part of the public transportation system in many major cities across the globe as they transport passengers in and out of stations as well as between the concourse and the tracks. 

MTR, the major Hong Kong public transport provider, has been operating a public transportation network consisting of heavy rail, light rail, and bus services centred around a 10-line subway network serving the urbanised areas of Hong Kong, since 1975. The average daily ridership is approximately 4.5 million patronages \citep{MTRHKPatronage}, which accounts for approximately 48.3\% (in 2022) of the total Hong Kong public transportation patronage \citep{MTRHKTrans}. 

The MTR operates more than 1000 escalators, these escalators are installed in various railway lines and stations with different ages, vertical rises, and workloads. Escalators in the same station are usually of the same age. An escalator's refurbishment is usually linked with its design life as recommended by the manufacturer, typically 25 to 35 years based on the industrial practice. However, the actual useful life of an escalator should be determined by its operating condition which is affected by the runtime, workload, maintenance quality, vibration, etc., rather than age only. Under the ``time-based'' strategy, all escalators in a station need to be refurbished more or less at the same time and the stoppage may be up to 3 months. This will inevitably cause inconvenience to the passengers and hence affect the level of service. If the refurbishment work is postponed without fully understanding the health condition of the escalators, the escalators may not operate well.

An alternative strategy is the use of data analytics models to predict remaining useful life (RUL) to inform refurbishment planning of escalators in accordance with their operating condition, rather than their age only. This predictive approach has been popular in the literature ever since the use of sensors and technology has become ubiquitous. \cite{lee2014prognostics} provided a comprehensive review of the systematic design of the PHM methodology, including procedures for identifying critical components, as well as tools for selecting the most appropriate algorithms for specific applications. \shortcites{lai2019internet}\citet{lai2019internet} demonstrated the benefits and challenges of condition-based maintenance (CBM) via the Internet of Things (IoT) relative to conventional corrective maintenance (CM) and time-based maintenance (TBM) in the case of elevators. \shortcites{awatramani2022investigating}\citet{awatramani2022investigating} investigated predicting the maintenance of an open-sourced elevator system using random forests. As for RUL prediction, \cite{li2020multi} proposed a multi-sensor data-driven RUL prediction method for semi-observable systems, which is based on a generalizable state-space model combined with a particle filtering framework. \cite{zhang2023health} characterized the health status of escalator rolling bears and performed RUL prediction using variational mode decomposition (VMD) and Gate Recurrent Unit (GRU) network. 

In this project, we develop a comprehensive health condition analytic system (HCAS) for escalators to support asset replacement planning and decisions. The system will use real-time sensor data combined with an analytic model. The system consists of four parts: 1) a data processing pipeline and database; 2) a dashboard for condition monitoring; 3) a health index model; and 4) an RUL prediction model. The system can be used for identifying escalators in poor operating conditions, in order to initialize relevant preventive maintenance measures, enhance passenger safety, and extend its operating lifetime. The system can also be used to identify those escalators in good condition in order to extend their lifetime before refurbishment so as to reduce wastage, carbon emission, etc. 

This paper is organised as follows. The next section provides an overview of the framework of the proposed HCAS. Section 3 discusses the data collection system and provides some descriptive data analysis. Section 4 introduces the first two parts of the HCAS: the data pre-processing and monitoring. Two kernel models of the HCAS are introduced in Sections 5 and 6, respectively. In Section 7, we summarize and discuss future work. 

\section{Project Overview}

The overarching project objective is to establish an Escalator Health Condition Analytic System (HCAS). To facilitate the development of the HCAS, sensors, and meters will be installed in a selection of escalators with different ages and operating conditions to collect the required data for analysis. See Section \ref{section:vibra_data_collection} for details on the selected types of data to be collected. 

\begin{figure}[h!t]
\centering
\includegraphics[width=0.9\textwidth]{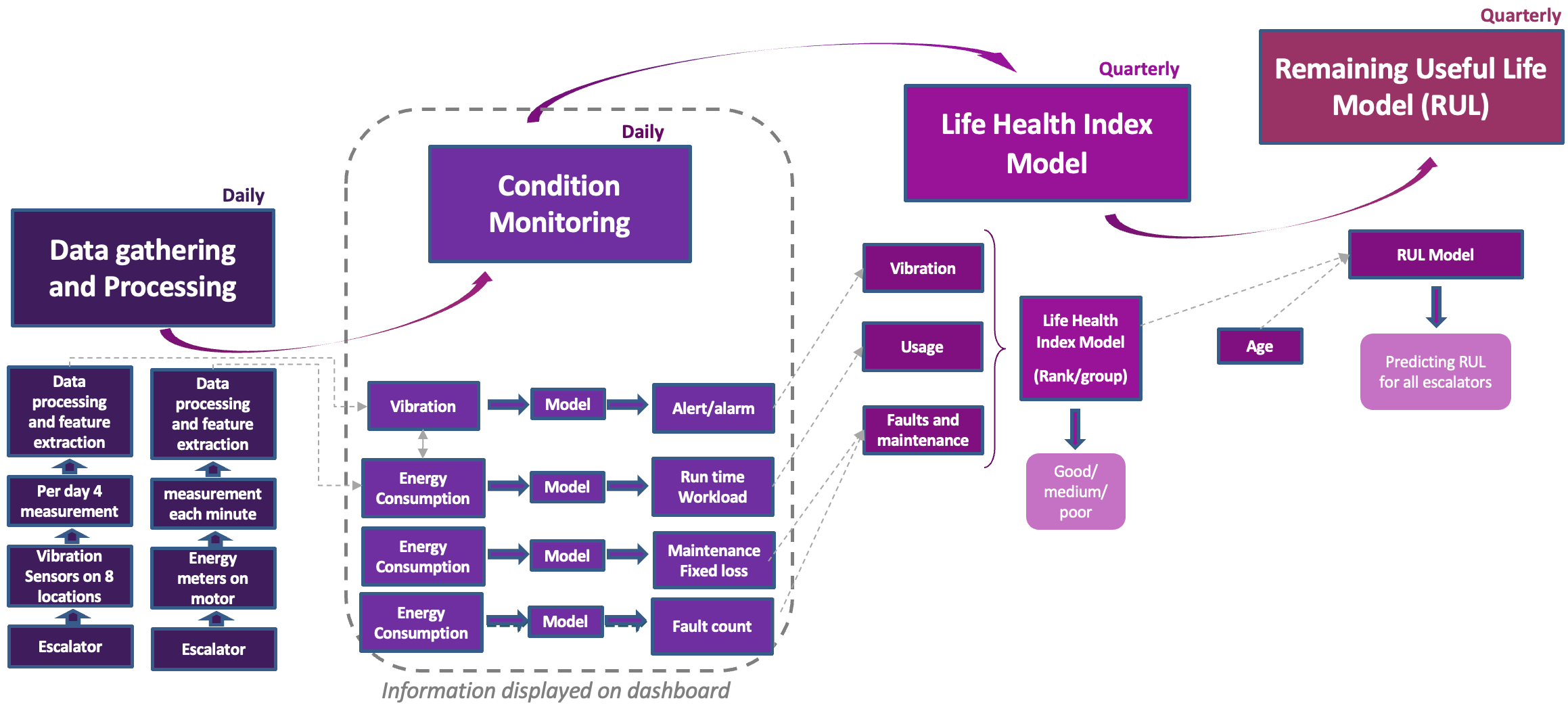}
\caption{\label{fig1:developed_framewok}Developed framework for the HCAS.}
\end{figure}

The proposed HCAS has an objective to estimate the escalator's remaining useful life (RUL) using the lifetime health index (LHI). The framework consists of four parts, each detailed in Figure \ref{fig1:developed_framewok}. The first part, data pre-processing, introduces the used data pre-processing and storing. ``Condition Monitoring'' is the second part focusing on feature engineering and the monitoring of escalators whose results are visualized on the dashboard. The third part develops the Health Index Model based on the lifetime health index (LHI). The fourth part is the remaining useful life model. Each of these parts will be discussed in Sections 4 to 6, respectively.

The developed HCAS can benefit escalator owners for better asset planning, enhanced safety, reduction of waste due to early refurbishment, and reduction of energy usage due to better insight into operating conditions and associated maintenance activities. Therefore, the HCAS can benefit from better utilization of assets and a reduction of operational expenditure.

For the development of the HCAS, various data sources are included. First of all, vibration, which is known to be a good indicator to reflect the condition of an escalator, especially of the individual components of the driving system. However as far as the lifetime prediction of a complete escalator is concerned, other critical parameters need to be considered. Lifetime of an escalator generally links with the years of operation. The design life recommended by the escalator manufacturer is based on the duty requirements. In reality, the escalator loading conditions and operating time vary for escalators installed at different locations in a station. As such it is found that some escalators which have lighter average workload and/or less average operating time, are still in good condition even when their designed lifetime has been reached. As the energy consumption of an escalator directly reflects its loading condition and the period of operation, it is also included as an important data source and input into the model. 

The higher the energy consumption the higher the expected wear-and-tear of the major components.  The phenomenon has been proved in a proof-of-concept study. In 2018, MTR engaged with City University of Hong Kong to carry out a Proof of Concept study to understand the correlation between energy consumption (workload) and step chain/main drive chain elongation for 20 escalators. The result was affirmative to show their positive relationship and it was proved that the workload was considered one of the factors which affect the condition of an escalator. Therefore, a further study on evaluation of the remaining useful life of an escalator by consideration of workload, the vibration of major components, and other parameters of the escalator was proposed.

Besides vibration and energy data, the failure history of an escalator also needs to be considered in order to establish an assessment of the impact on the escalator condition due to the frequency of incidents and their severity. Hence, the HCAS developed in this project is based on three input data sources: energy consumption, vibration data, and failure frequency.

\section{Data collection system and descriptive data analysis}

There are 12 stations in this project, each of which contains two test escalators. Detailed information, i.e., the vertical rise, running direction, and age, of these 24 escalators are given in Table \ref{table:detailed_infor_escalators}, where we can see most of the escalators are in the middle of their useful life, and there are also young (Escalator 0 and 1) and old escalators (Escalator 2 and 3) in the dataset.

\begin{table}[h!]
 \centering
\resizebox{\columnwidth}{!}{\begin{tabular}{cccc | cccc} 
 \hline \hline
 Escalator ID & Rise (m) & Direction & Age (Year) & Escalator ID & Rise (m) & Direction & Age (Year) \\ [0.5ex] 
 \hline
0 & 16.72 & Up & 7 & 12 & 3.567 & Up & 18.7 \\
1 & 16.72 & Up & 7 & 13 & 3.567 & Down & 18.7 \\
2 & 5.5 & Up & 24.2 & 14 & 6.79 & Up & 18.7 \\
3 & 5.5 & Down & 24.2 & 15 & 6.79 & Down & 18.7 \\
4 & 8.926 & Down & 18.7 & 16 & 5.35 & Bi-Dire & 17.7 \\
5 & 8.926 & Down & 18.7 & 17 & 5.35 & Bi-Dire & 17.7 \\
6 & 5.6 & Down & 13.1 & 18 & 8.175 & Up & 18.7 \\
7 & 5.6 & Up & 13.1 & 19 & 8.175 & Down & 18.7 \\
8 & 8.28 & Down & 18.7 & 20 & 6.02 & Down & 18.7 \\
9 & 8.28 & Down & 18.7 & 21 & 6.02 & Up & 18.7 \\
10 & 5.88 & Down & 18.7 & 22 & 7.635 & Up & 18.7 \\
11 & 5.88 & Up & 18.7 & 23 & 7.635 & Down & 18.7 \\
 \hline \hline
\end{tabular}}
\caption{Detailed information of the escalators.}
\label{table:detailed_infor_escalators}
\end{table}

\begin{figure}[h!t]
\centering
\includegraphics[width=0.9\textwidth]{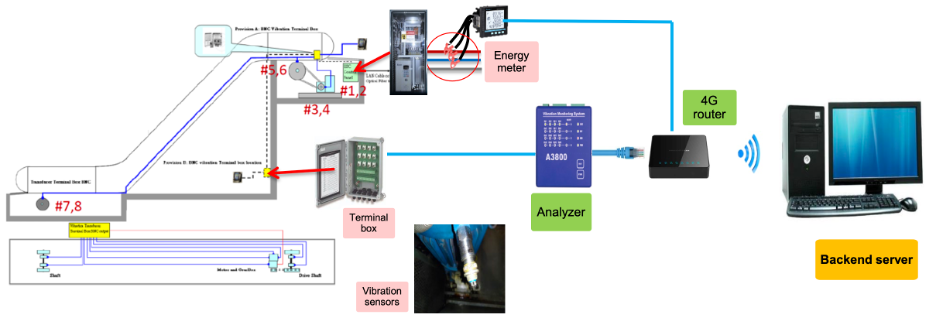}
\caption{\label{fig:framework_DCS}The framework of the online data collection system.}
\end{figure}

To collect the online data, a data collection system is installed in each of the escalators, where both the energy data and vibration data are collected. The framework of the system is shown in Figure \ref{fig:framework_DCS}, where the energy meter and the vibration sensors are installed in the escalator. These data are then transmitted by a 4G router to a backend server, on which they are stored for further analysis.

\subsection{Online energy collection}

The energy meter records the whole day's data of an escalator with an 1-minute sample interval. These data include the recording time, current, voltage, and power consumption values. Based on these data, the total energy consumption value per minute of an escalator can be computed as Eq. \eqref{eq:total_energy_consump}

\begin{equation}
E_{T} = E_{Imp} + E_{Exp}, \label{eq:total_energy_consump}
\end{equation}

\noindent where \(E_{T}\)  is the total energy consumption value, and \(E_{Imp}\) and \(E_{Exp}\) are the instantaneous power input and the instantaneous regenerative power, respectively.

\begin{figure}
     \centering
     \begin{subfigure}[b]{0.48\textwidth}
         \centering
         \includegraphics[width=\textwidth]{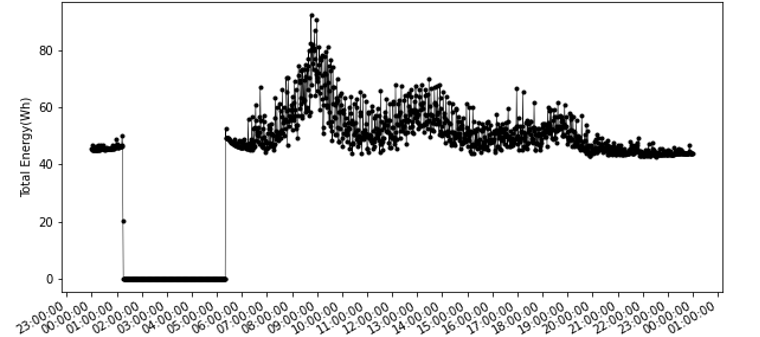}
         \caption{An upward escalator}
         \label{fig:total_energy_consump_upward}
     \end{subfigure}
     \hfill
     \begin{subfigure}[b]{0.48\textwidth}
         \centering
         \includegraphics[width=\textwidth]{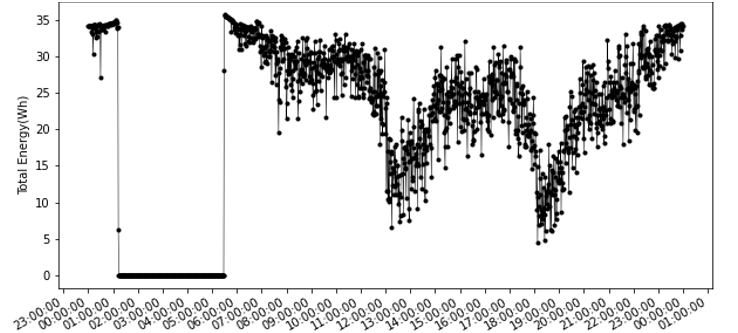}
         \caption{A downward escalator}
         \label{fig:total_energy_consump_downward}
     \end{subfigure}
        \caption{The total energy consumption values through time.}
        \label{fig:total_energy_consump}
\end{figure}

The daily total energy consumption plots are useful for understanding the working patterns of the escalator. These plots are different for upward and downward escalators due to the earth's gravity. Figure \ref{fig:total_energy_consump} shows the total energy consumption values versus time on one day from escalators in an upward and a downward direction. 

Compared with Figures \ref{fig:total_energy_consump_upward} and \ref{fig:total_energy_consump_downward}, the total energy consumption value goes up for an upward escalator, especially during rush hours (9:00, 14:00, and 19:00), while the total energy consumption value drops dramatically for a downward escalator during these hours. The reason behind this is that the upward escalator needs more energy to lift the passengers, while for the downward escalator, gravity will drive the passengers to go down with surplus energy being produced, and therefore save the energy consumption value of the escalator.

\subsection{Online vibration data collection} \label{section:vibra_data_collection}

Apart from the energy data, the data collection system also collects the vibration data of the escalator. As seen in Figure \ref{fig:framework_DCS}, 8 vibration sensors are installed, and their locations are labelled in red numbers. Table \ref{table:descrip_vibra_sensors} gives a detailed description of these vibration sensors, which are related to 4 major components (both driven end (DE) and non-driven end (NDE)) of the escalator. Specifically, the gearbox and motor are high-frequency components as they transmit the power of the escalator directly from the engine, while the main drive and tension carriage bearings are far from the engine and therefore are low-frequency components.

\begin{table}[h!]
\centering
\begin{tabular}{ccc} 
 \hline \hline
Point ID & Vibration Sensor Location & Direction \\ [0.5ex] 
 \hline
1 & Gearbox bearing DE & Horizontal \\
2 & Gearbox bearing NDE & Vertical \\
3 & Moter bearing DE & Axial \\
4 & Motor bearing NDE & Loading side \\
5 & Main shaft bearing DE & Loading side \\
6 & Main shaft bearing NDE & Loading side \\
7 & Tension carriage bearing DE & Loading side \\
8 & Tension carriage bearing NDE & Loading side \\
\hline \hline
\end{tabular}
\caption{Description of vibration sensors.}
\label{table:descrip_vibra_sensors}
\end{table}

The vibration sensor is able to collect different types of measurements of the major component. For this project we use the acceleration data in the frequency domain (fft\_g). Due to the capacity limit, the original time-domain spectrum data could not be stored directly, it is edge-processed to reduce its data size. Fast Fourier transform (FFT) algorithm is applied to the acceleration data.

\section{Data pipeline: data preparation, database and dashboard} \label{section:data_pipeline}

After the raw energy and vibration data are gathered, some data cleaning and feature extraction techniques are applied. The details of how to process raw data are discussed in the first two parts. Both raw and processed data are stored in a database, and the database structure is introduced in the third part of this subsection. The processed data are used to monitor the operating condition of escalators, which are displayed on a dashboard. The layout of the dashboard is introduced in the last subsection.

\subsection{Vibration data pre-processing}

The raw vibration data may contain unexpected noise in certain frequency bands, which should be processed first. And then, the key feature (i.e., \(A_{t}\)) is computed to show the summary energy level for the vibration data. A detailed discussion of these two procedures will be presented in the following sub-section.

\subsubsection{Dominant frequency band selection}

To reduce the noise and the data size for the raw vibration data, we first select the frequency bands with consistently high energy levels across all escalators. As suggested by the expert, the on-site noise has larger impact on the low frequency bands, while the high frequency bands (usually larger than 10 kHz) contain low energy level, therefore these frequency bands can be safely removed in data pre-processing. The detailed steps for choosing the dominant frequency bands for each sensor are provided as below.

\begin{enumerate}
	\item Group the vibration data according to different sensors,
	\item Divide the frequency range (0 \(\sim\) 12.8 kHz) into frequency bands (per 1 kHz),
	\item Compute the RMS values of each frequency band,
	\item Make bar plots and select the frequency bands with higher energy levels and fewer extreme values.
\end{enumerate}

The RMS represents the energy level for each frequency band and is defined as follows,

\begin{equation}
	RMS = \sqrt{\frac{\sum_{i=s}^{m}x_{i}^{2}}{m-s+1}},  \label{eq:rms_defin}
\end{equation}

\noindent where \(s\) and \(m\) are the starting and ending indices of this frequency band, \(x_{i}\) is the \(i\)th component of the vibration data, and \(m-s+1\) is the counts of vibration data within this frequency band. 

\begin{figure}[h!t]
	\centering
	\begin{subfigure}[b]{0.475\textwidth}
		\centering
		\includegraphics[width=\textwidth]{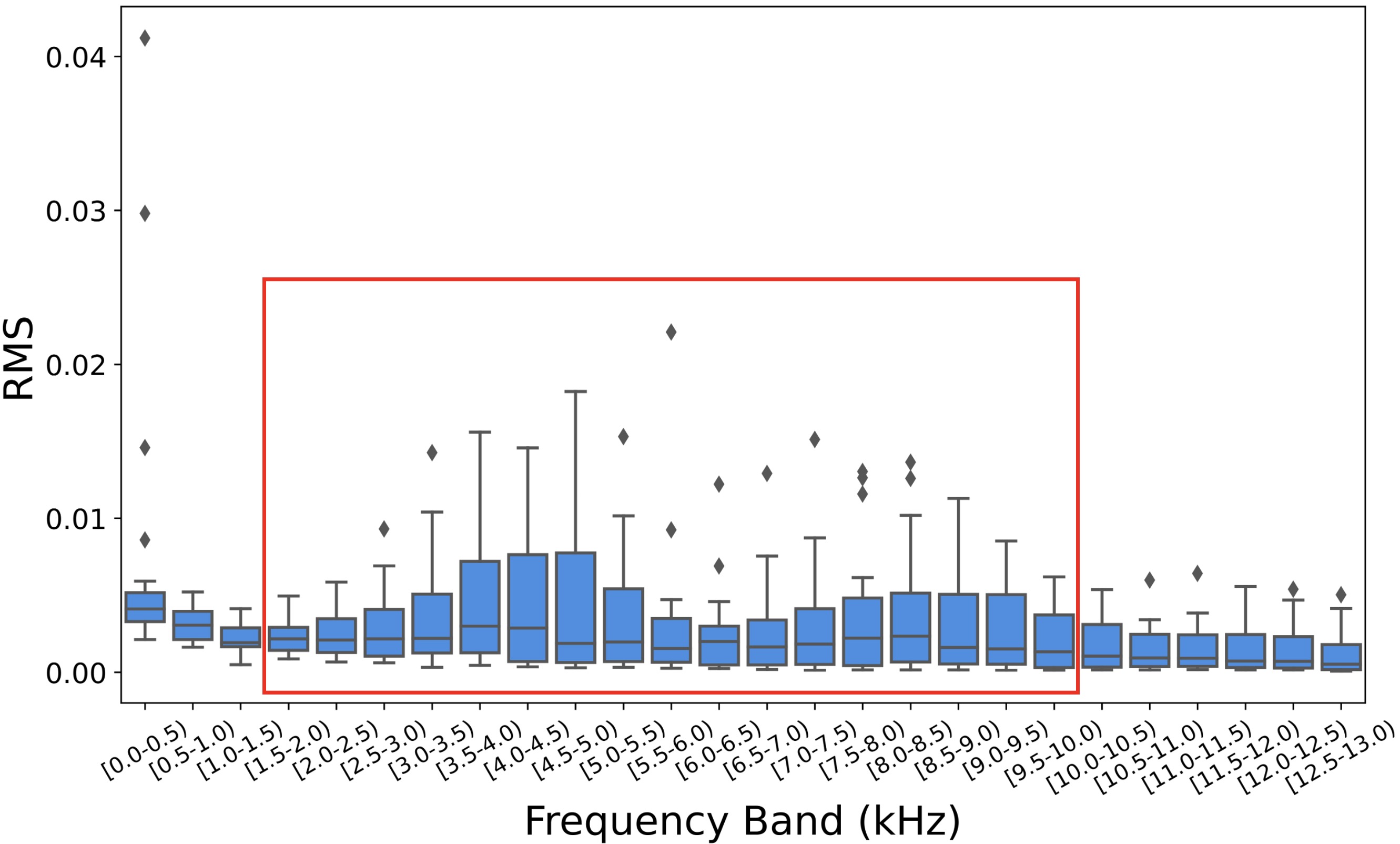}
		\caption[]%
		{{\small Gearbox DE}}    
		\label{fig:mean and std of net14}
	\end{subfigure}
	\hfill
	\begin{subfigure}[b]{0.475\textwidth}  
		\centering 
		\includegraphics[width=\textwidth]{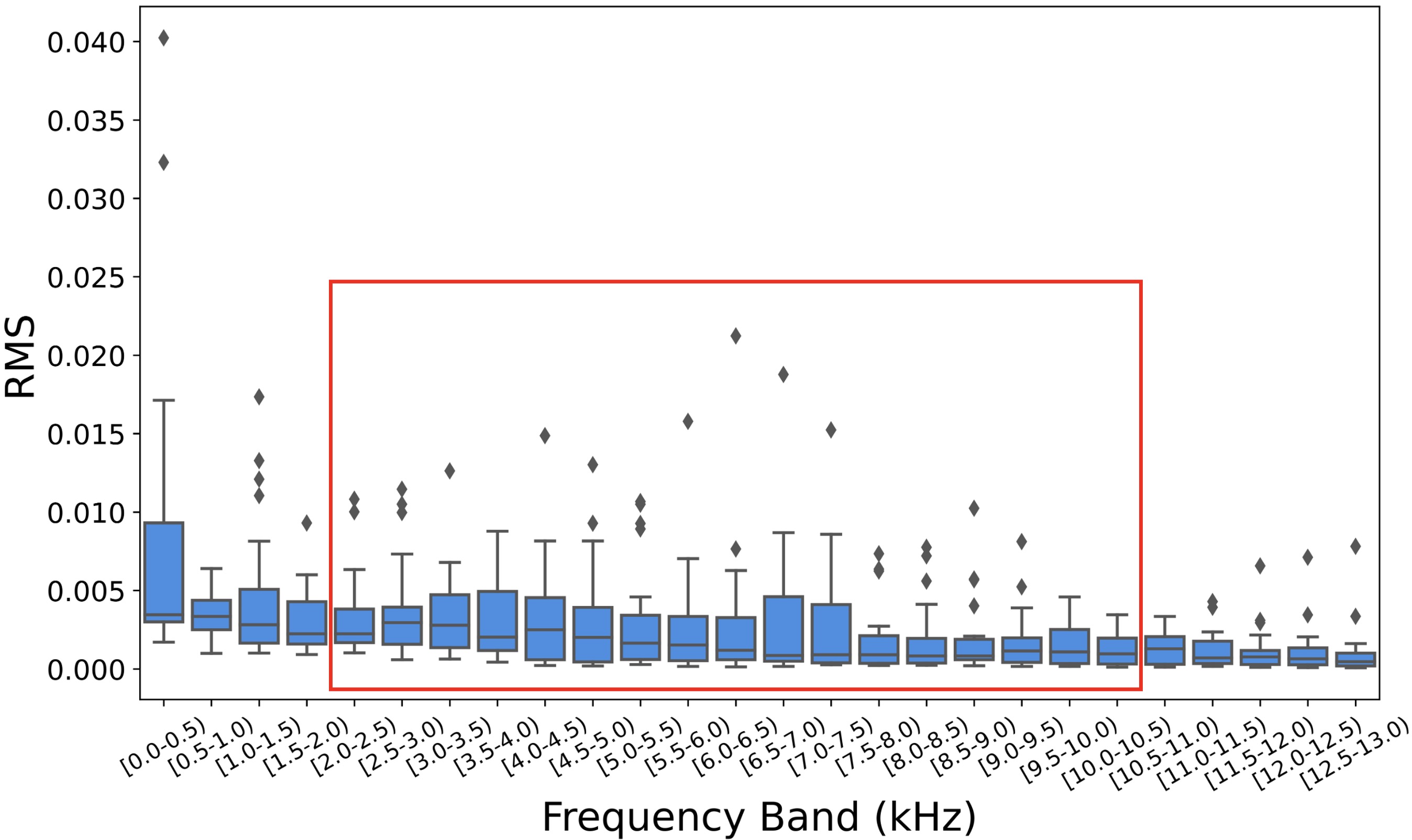}
		\caption[]%
		{{\small Gearbox NDE}}    
		\label{fig:mean and std of net24}
	\end{subfigure}
	\vskip\baselineskip
	\begin{subfigure}[b]{0.475\textwidth}   
		\centering 
		\includegraphics[width=\textwidth]{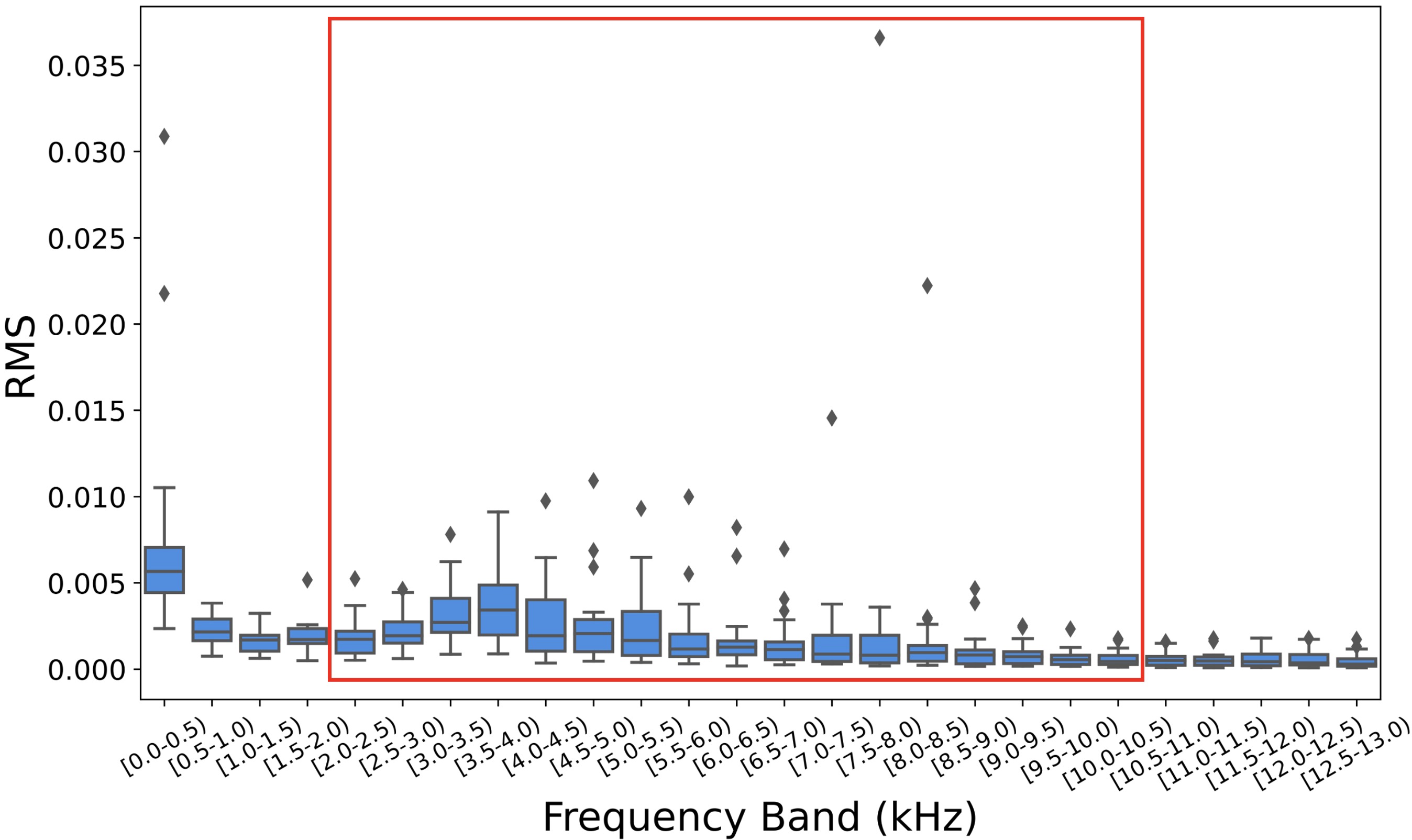}
		\caption[]%
		{{\small Motor DE}}    
		\label{fig:mean and std of net34}
	\end{subfigure}
	\hfill
	\begin{subfigure}[b]{0.475\textwidth}   
		\centering 
		\includegraphics[width=\textwidth]{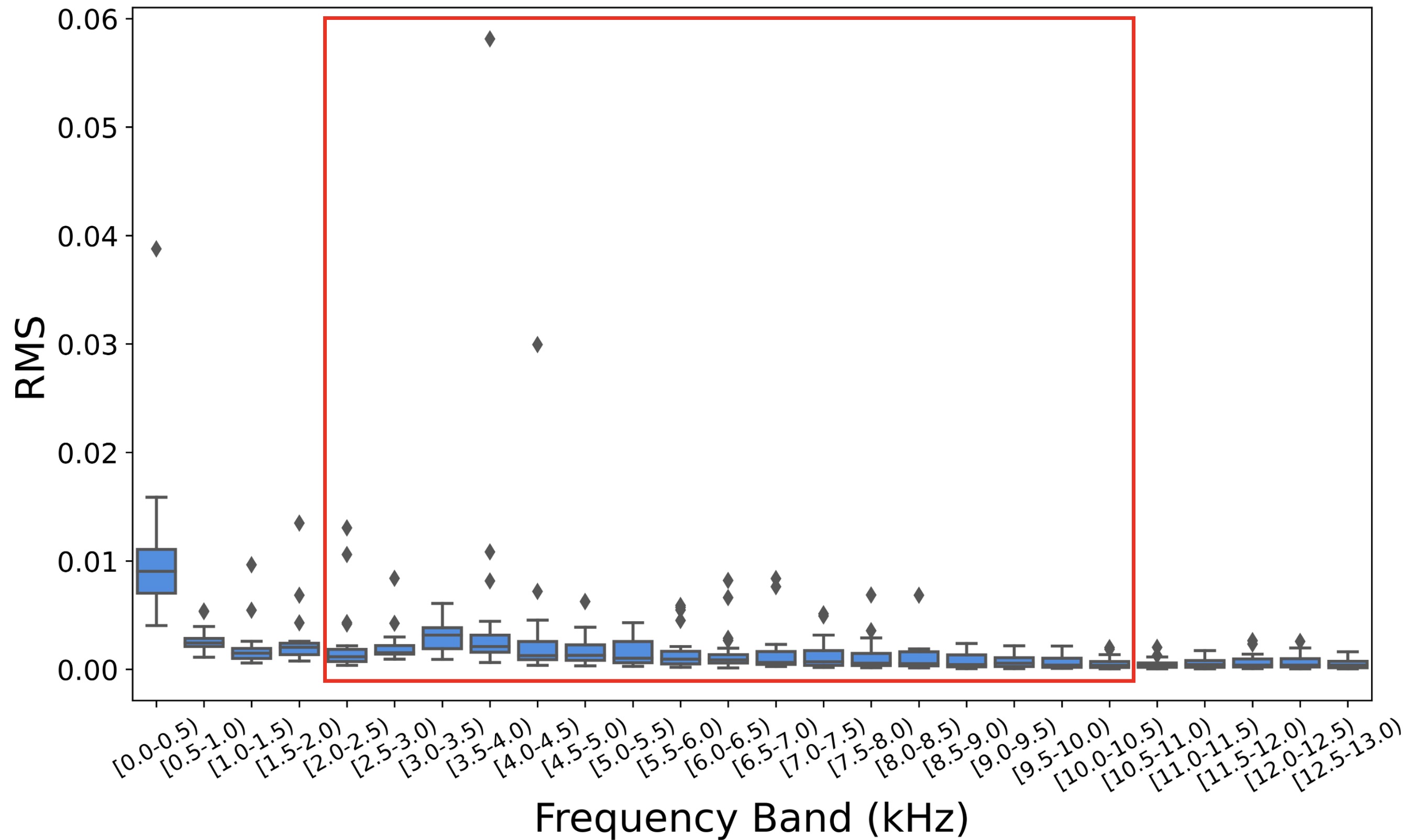}
		\caption[]%
		{{\small Motor NDE}}    
		\label{fig:mean and std of net44}
	\end{subfigure}
	\caption[Frequency band selection for high-frequency components.]
	{\small Frequency band selection for high-frequency components.} 
	\label{fig:freq_sel_h}
\end{figure}

\begin{figure}[h!t]
	\centering
	\begin{subfigure}[b]{0.475\textwidth}
		\centering
		\includegraphics[width=\textwidth]{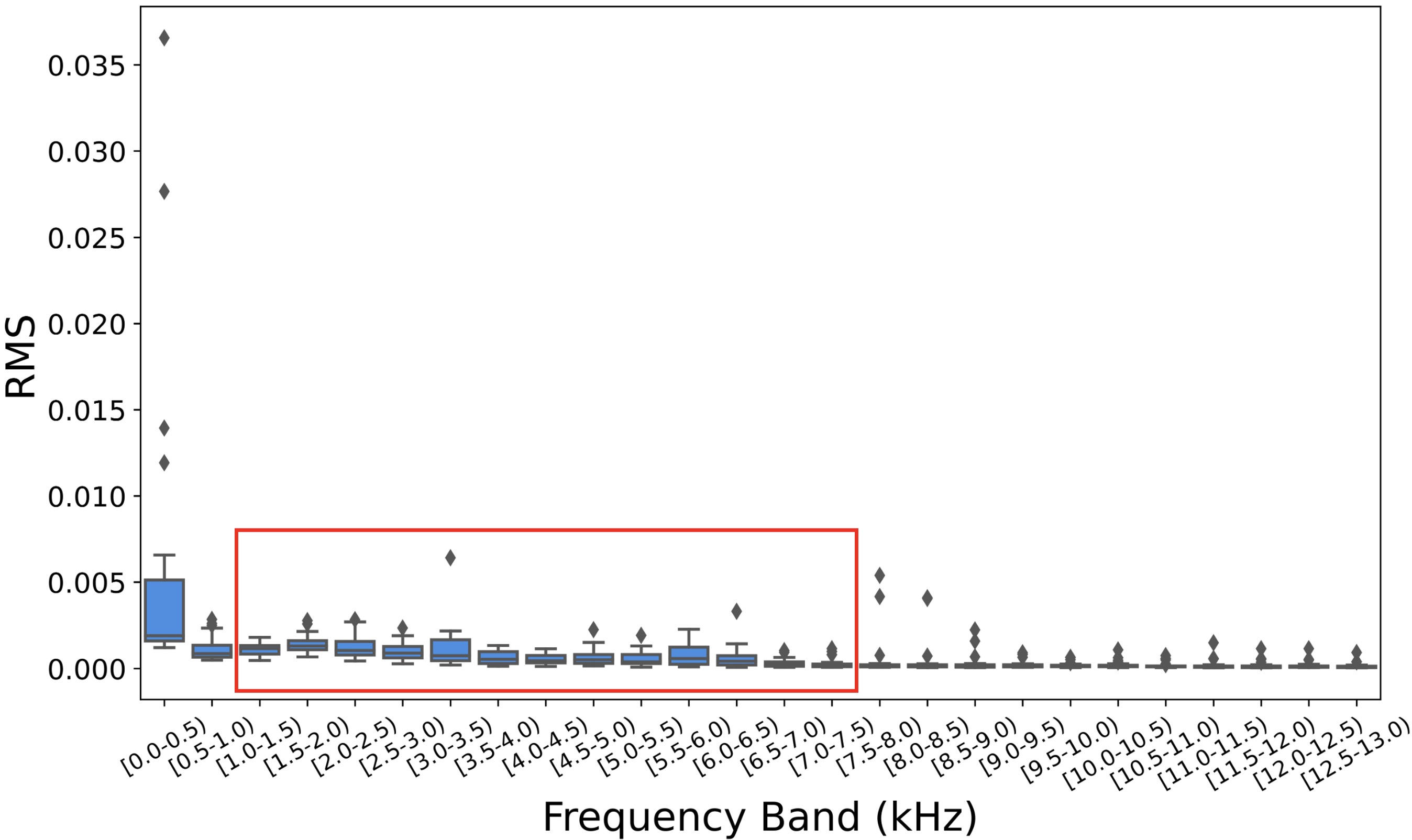}
		\caption[]%
		{{\small Main Drive DE}}    
		\label{fig:mean and std of net15}
	\end{subfigure}
	\hfill
	\begin{subfigure}[b]{0.475\textwidth}  
		\centering 
		\includegraphics[width=\textwidth]{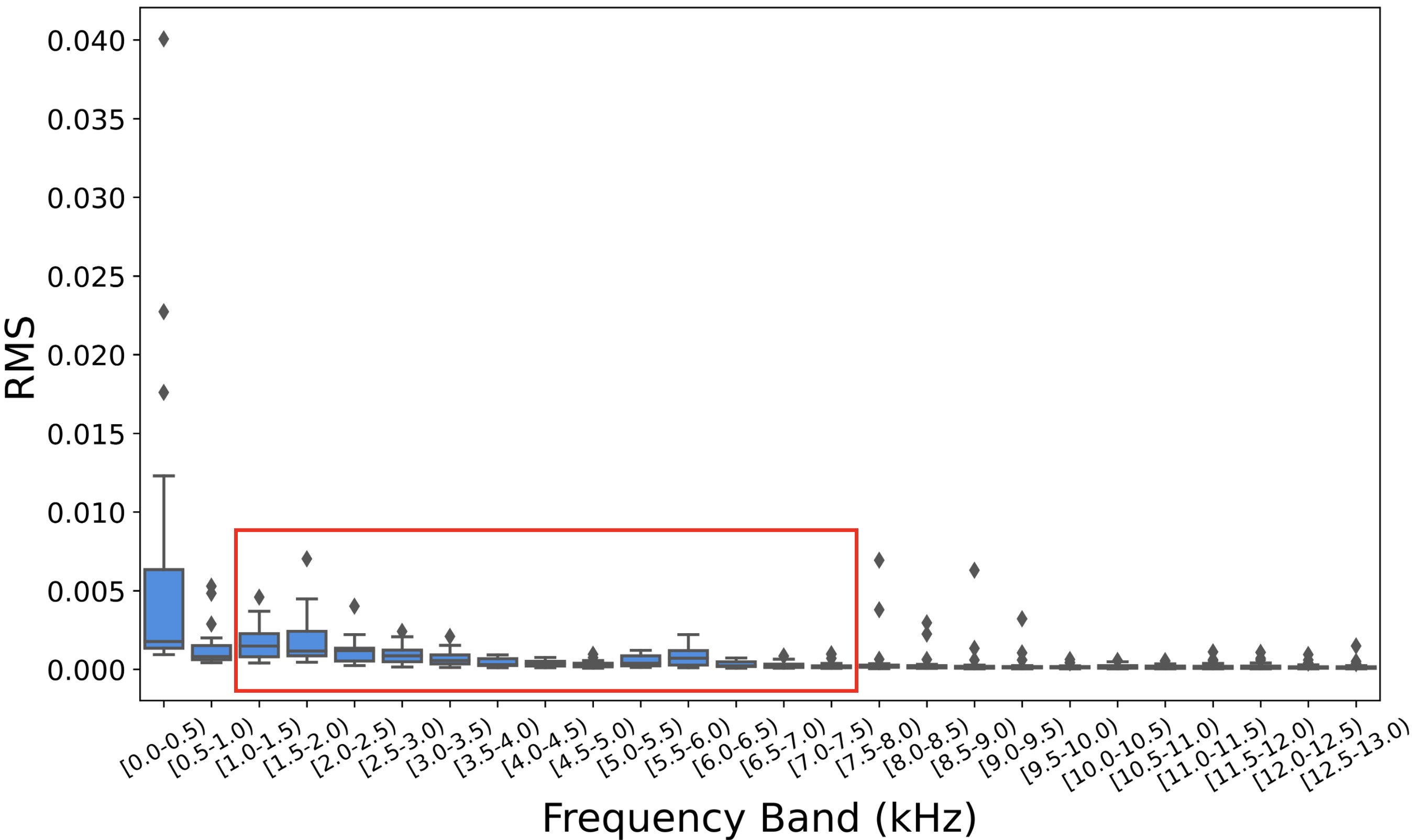}
		\caption[]%
		{{\small Main Drive NDE}}    
		\label{fig:mean and std of net25}
	\end{subfigure}
	\vskip\baselineskip
	\begin{subfigure}[b]{0.475\textwidth}   
		\centering 
		\includegraphics[width=\textwidth]{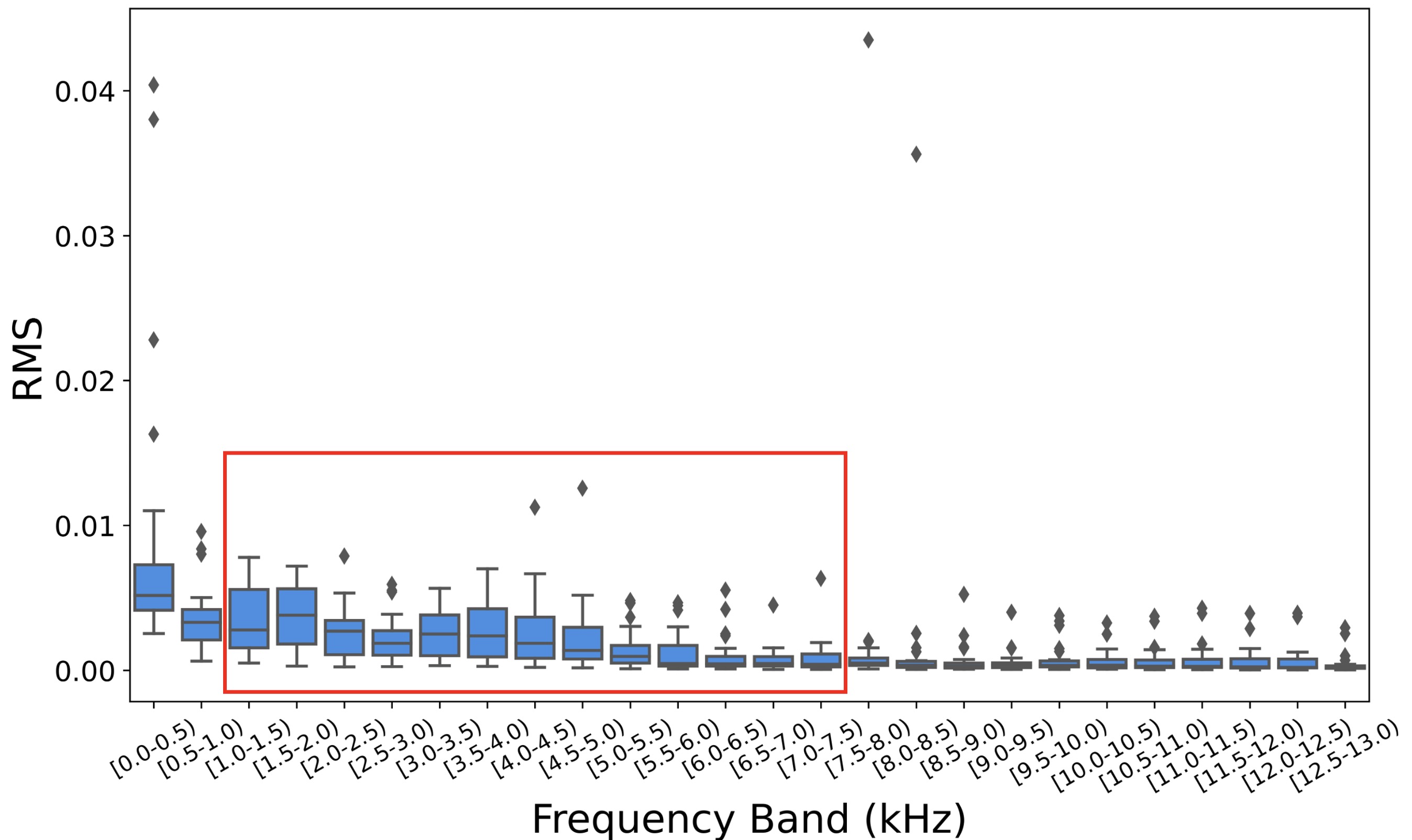}
		\caption[]%
		{{\small Tension carriage left DE}}    
		\label{fig:mean and std of net35}
	\end{subfigure}
	\hfill
	\begin{subfigure}[b]{0.475\textwidth}   
		\centering 
		\includegraphics[width=\textwidth]{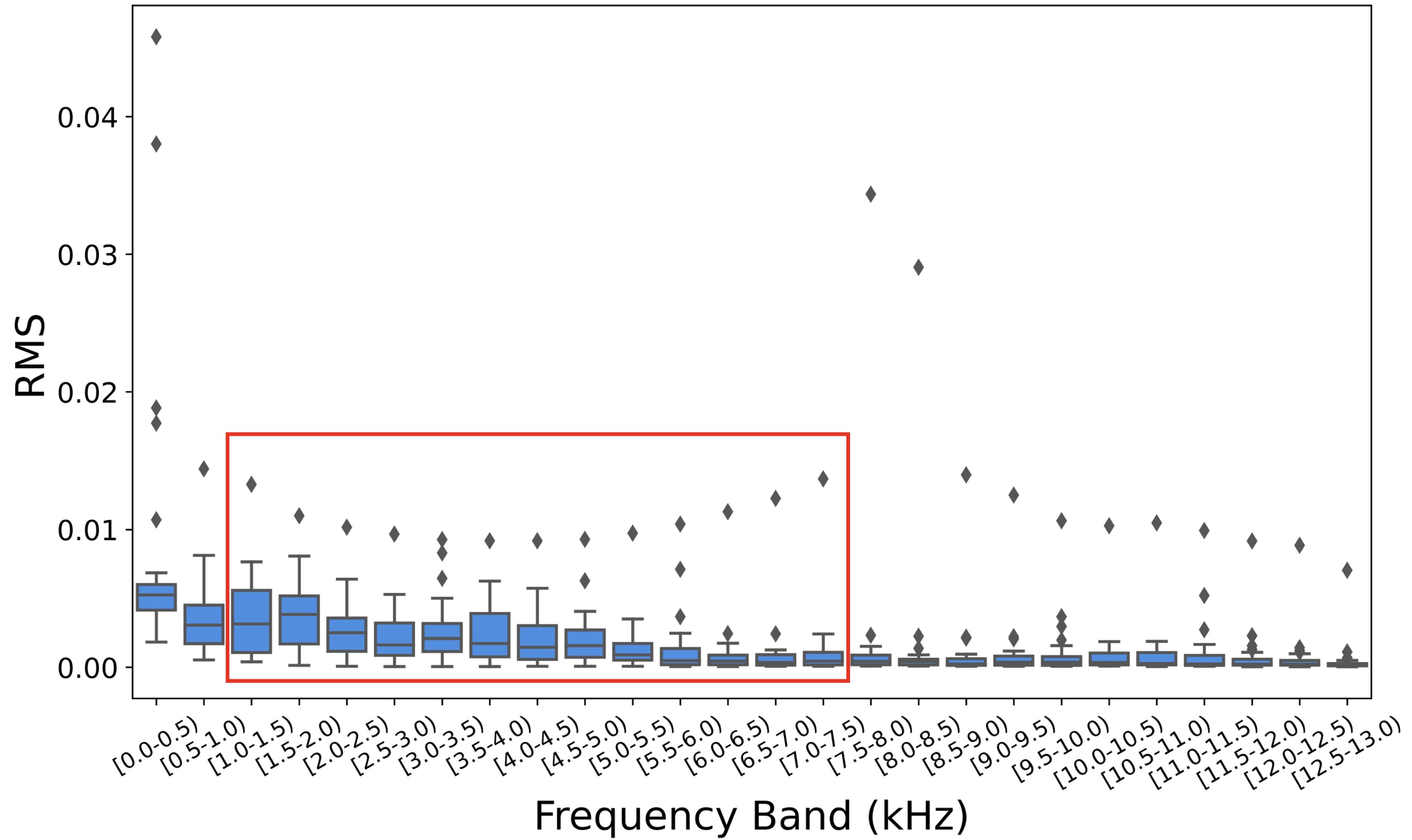}
		\caption[]%
		{{\small Tension carriage right NDE}}    
		\label{fig:mean and std of net45}
	\end{subfigure}
	\caption[Frequency band selection for low-frequency components.]
	{\small Frequency band selection for low-frequency components.} 
	\label{fig:freq_sel_l}
\end{figure}

Based on the distinct operational mechanisms of each component, we have categorized them into two groups: high-frequency and low-frequency components. Specifically, given that the gearbox and motor typically operate at higher frequencies in comparison to the main drive and tension carriage, we have classified the former as high-frequency components and the latter as low-frequency components. Then, box plots are made to help understand the process of choosing dominant frequency bands for high-frequency and low-frequency bands (see Figures \ref{fig:freq_sel_h} and \ref{fig:freq_sel_l}).

Figure \ref{fig:freq_sel_h} shows the frequency bands for the high-frequency components, revealing that the [0 \(\sim\) 0.5] kHz frequency band features numerous extreme values, indicative of significant noise. In contrast, the [0.5 \(\sim\) 2] kHz and [10 \(\sim\) 13] kHz frequency bands exhibit relatively low energy levels and can be safely disregarded with minimal loss of information, as shown in Figure \ref{fig:freq_sel_l}. Based on these findings, we selected the [2 \(\sim\) 10] kHz frequency band for the high-frequency components. Similarly, we selected the [1 \(\sim\) 7.5] kHz frequency band for the low-frequency components, following the same rationale (see Table \ref{table:category_dfb}). Henceforth in this paper, unless otherwise specified, the term ``vibration data'' specifically refers to the data that has gone through frequency band selection.

\begin{table}[h!t]
	\centering
	\begin{tabular}{c  c  c} 
		\hline \hline
		Category & Component & Dominant Frequency Bands \\ [0.2em]
		\hline
		High-frequency components & Gearbox and Motor & [2 \(\sim\) 10] kHz \\
		Low-frequency components & Main Drive and Tension Carriage & [1 \(\sim\)   7.5] kHz \\
		\hline \hline
	\end{tabular}
	\caption{Components categories and dominant frequency bands.}
	\label{table:category_dfb}
\end{table}

\subsubsection{\(A_{t}\) values}

We process the daily data to 3 records per day (some raw vibration sensor data is collected out of working time). The \(A_{t}\) values show the vibration level of a component, and its value is computed as:

\begin{equation}
    A_{t} = \sqrt{\frac{\sum_{i}x_{i}^{2}}{1.5}}, \label{eq:at_equa}
\end{equation}

\noindent where \(x_{i}\) is the \(i\)th component of the vibration data \(x\), and teh factor 1.5 is suggested by the MTR vibration expert.

If the $A_{t}$ is larger than a preset threshold, the alert or alarm should be issued. The alert and alarm thresholds for different sensors are shown in Table \ref{table:at_thresholds}.

\begin{table}[h!]
\centering
\begin{tabular}{c|c c c c} 
 \hline \hline
\multirow{3}{*}{Sensor Name} & \multicolumn{4}{c}{\(A_{t}\) Thresholds} \\ [0.2ex] 
& \multicolumn{2}{c}{Acceleration (g)} & \multicolumn{2}{c}{Velocity (mms)} \\ [0.2ex] 
& Alert & Alarm & Alert & Alarm \\ [0.5ex] 
\hline
Gearbox DE & 0.375 & 0.75 & 2.8 & 4.5 \\ [0.2ex] 
Gearbox NDE & 0.375 & 0.75 & 2.8 & 4.5 \\ [0.2ex] 
Main Drive DE & 0.15 & 0.3 & 2.8 & 4.5 \\ [0.2ex] 
Main Drive NDE & 0.15 & 0.3 & 2.8 & 4.5 \\ [0.2ex] 
Motor DE & 0.375 & 0.75 & 2.8 & 4.5 \\ [0.2ex] 
Motor NDE & 0.375 & 0.75 & 2.8 & 4.5 \\ [0.2ex] 
Tension carriage left DE & 0.15 & 0.3 & 2.8 & 4.5 \\ [0.2ex] 
Tension carriage right NDE & 0.15 & 0.3 & 2.8 & 4.5 \\ [0.5ex] 
\hline \hline
\end{tabular}
\caption{Preset \(A_{t}\) thresholds for vibration based on ISO standards.}
\label{table:at_thresholds}
\end{table}

\begin{figure}[h!]
\centering
\includegraphics[width=0.8\textwidth]{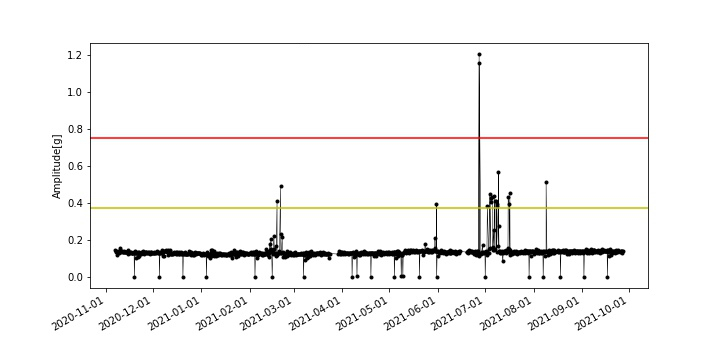}
\caption{\label{fig:acceleration_at}Acceleration \(A_{t}\) of Gearbox NDE of Escalator 7.}
\end{figure}

Hereunder is Figure \ref{fig:acceleration_at} demonstrating the \(A_{t}\) values of the FFT acceleration data of the gearbox NDE in Escalator 7. And the yellow and red lines are the alert and alarm thresholds, respectively. 

\subsubsection{$A_{t}$ exceedance curve area}
The preset thresholds given in Section 4.1.2 based on ISO standard work well on identifying extreme values. However, it is less useful in daily monitoring, as most escalators operate normally and no alarm will be triggered. Therefore a different  vibration statistic, the exceedance curve, is introduced.

The exceedance curve is a statistical tool representing the probability of exceeding specific thresholds of given variables. It is widely used in the fields, like risk assessment \cite{piechota2001development} . In our case, given a set of \(A_{t}\) values of sensor $i$, \(X_{i}\), the cumulative probability of a given threshold $\tau \in \left[0,\max(X_{i})\right]$ is computed by,
\begin{equation}\label{eq:exceedance_curve}
		p_{i,\tau} = \frac{\sum I\left(x > \tau \right)}{|X_{i}|},
\end{equation}
where \(I\) is the indicator function and \(|X_{i}|\) is the number of elements in \(X_{i}\).

The area of the exceedance curve is employed to indicate the vibration level of the escalator component sensor $i$, which is computed as follows.

\begin{equation}
	S_{i} = \int p_{i,\tau} \, d \tau .
\end{equation}

\begin{figure}
	\begin{subfigure}{0.49\linewidth}
		\includegraphics[width=\textwidth]{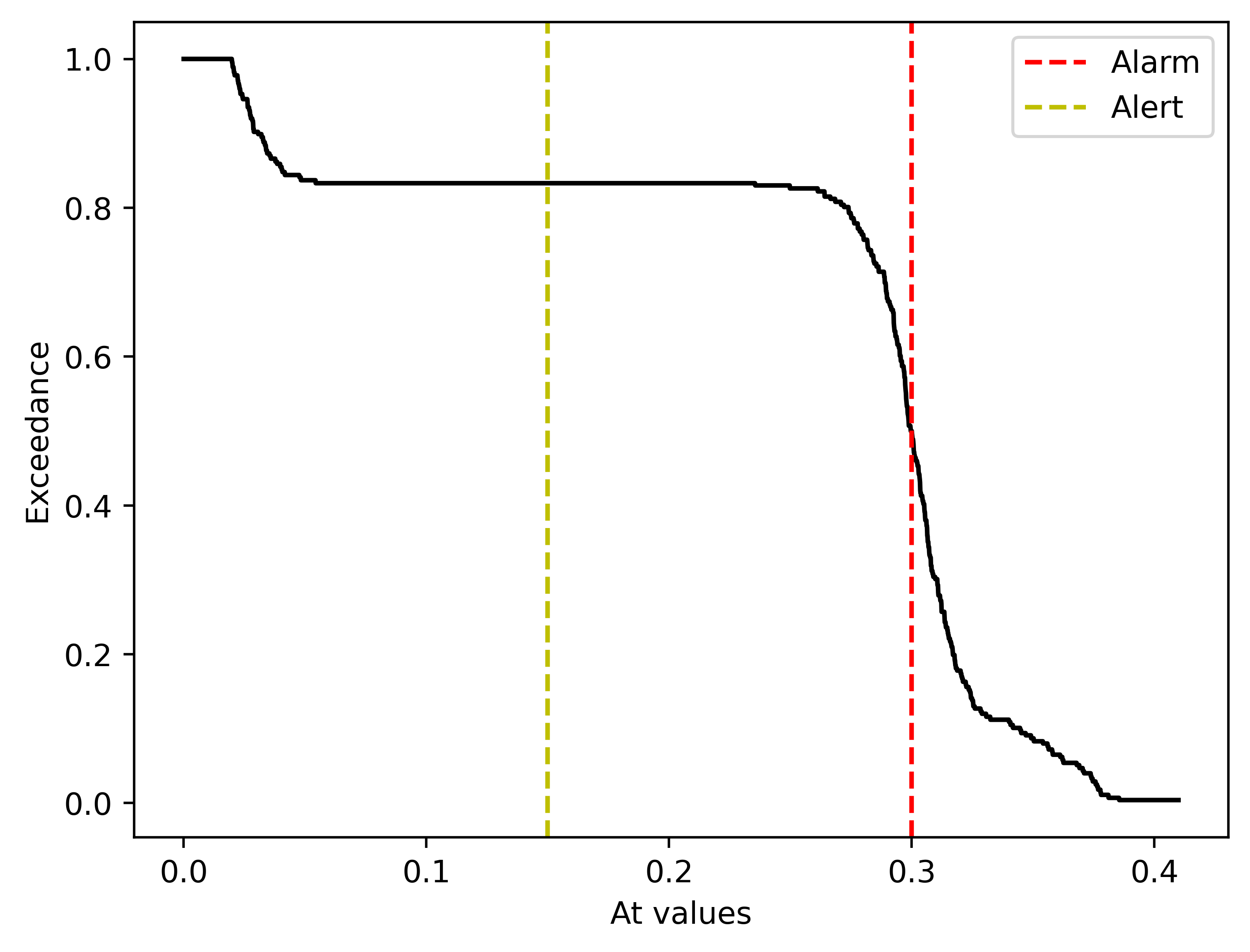}
		\caption{Escalator 0}
		\label{fig:ExCurve_alarms}
	\end{subfigure}
	\begin{subfigure}{0.49\linewidth}
		\includegraphics[width=\textwidth]{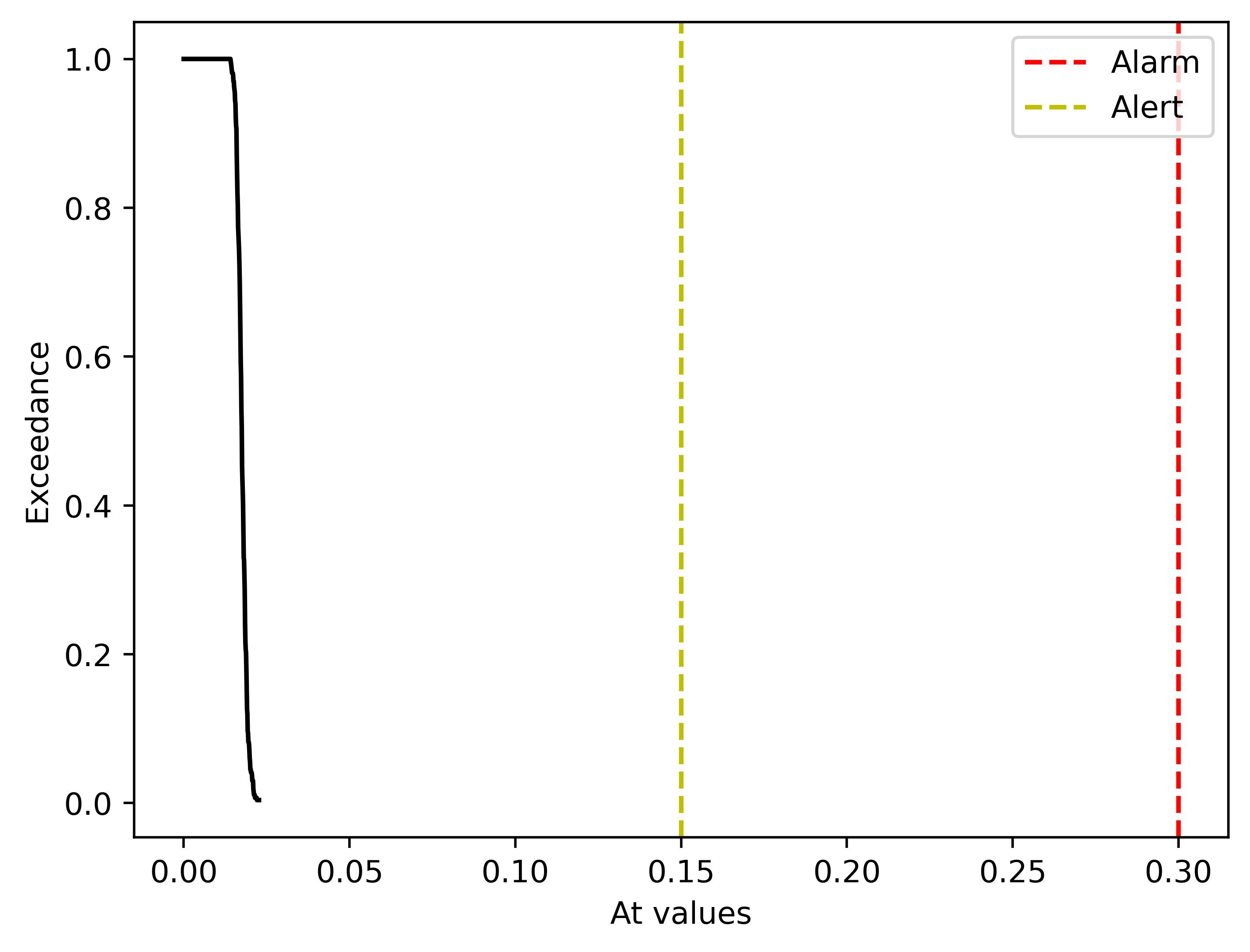}
		\caption{Escalator 3}
		\label{fig:ExCurve_noalarm}
	\end{subfigure}
	\caption{Exceedance Curve of Tension Carriage Left $A_{t}$ Values in 2021 Q4}
	\label{fig:ExCurveExample}
\end{figure}

Figure \ref{fig:ExCurveExample} present two exceedance curves of an escalator sensors' \(A_{t}\) values in 2021 Q4. Specifically, there are about $50\% A_{t}$ points of Tension Carriage Left of Escalator 0 exceeds the preset threshold $0.3$ as shown in Figure \ref{fig:ExCurve_alarms} indicating a worse vibration level, while all points of Escalator 3 are less than about $0.025$ showing a better vibration level. It is obvious that the curve with a larger area indicates a worse status of the component.

The vibration status of an escalator can be computed as the weighted sum of the areas of each sensor as shown in Eq. (\ref{eq:exceedance_curve_esc}).
\begin{equation}\label{eq:exceedance_curve_esc}
	N = \sum_{i}^{8}{W_{i} \cdot S_{i}},
\end{equation} 
where $W_{i}$ is given in the Table \ref{table:sensor_weight}.

\begin{table}[h!]
	\centering
	\begin{tabular}{cc } 
		\hline \hline
		Sensor Name & Weights $W_{i}$ \\ [0.5ex] 
		\hline
		Gearbox DE 					& 0.11 \\ [0.2ex] 
		Gearbox NDE 				& 0.11 \\ [0.2ex] 
		Main Drive (DE) 			& 0.17 \\ [0.2ex] 
		Main Drive (NDE) 			& 0.16 \\ [0.2ex] 
		Motor DE 					& 0.11 \\ [0.2ex] 
		Motor NDE 					& 0.11 \\ [0.2ex] 
		Tension carriage left (DE)  & 0.12 \\ [0.2ex] 
		Tension carriage right (NDE)& 0.11 \\ [0.5ex] 
		\hline \hline
	\end{tabular}
	\caption{Weights of vibration sensors.}
	\label{table:sensor_weight}
\end{table}

\subsection{Energy data processing}

The energy data are reorganized firstly by a service day, from 4 am to 4 am, instead of a calendar day. The daily energy consumption is recorded in minutes from the energy power meter readers. Through daily energy consumption, features like working time, fixed energy loss and passenger load can be extracted. And maintenance events can be identified from energy consumption by algorithms.

The fixed loss energy (\(E_{F}\)) is no-load energy consumption by minute, while the consumed energy with loads is named variable energy loss (\(E_{V}\)). Both fixed and variable energy loss is affected by the operating direction because of gravity. Specifically, the variable loss is always higher than the fixed loss for upward escalators, which is opposite to downward escalators. This relationship can be summarized by the following equation. 

\begin{equation}
    E_{T} = 
    \begin{cases}
      E_{F} + E_{V} & \text{when direction is up,}\\
      E_{F} - E_{V} & \text{when direction is down.}
    \end{cases} \label{eq:total_energy_fixed_variable}
\end{equation}

\begin{figure}[h!t]
\centering
\includegraphics[width=0.7\textwidth]{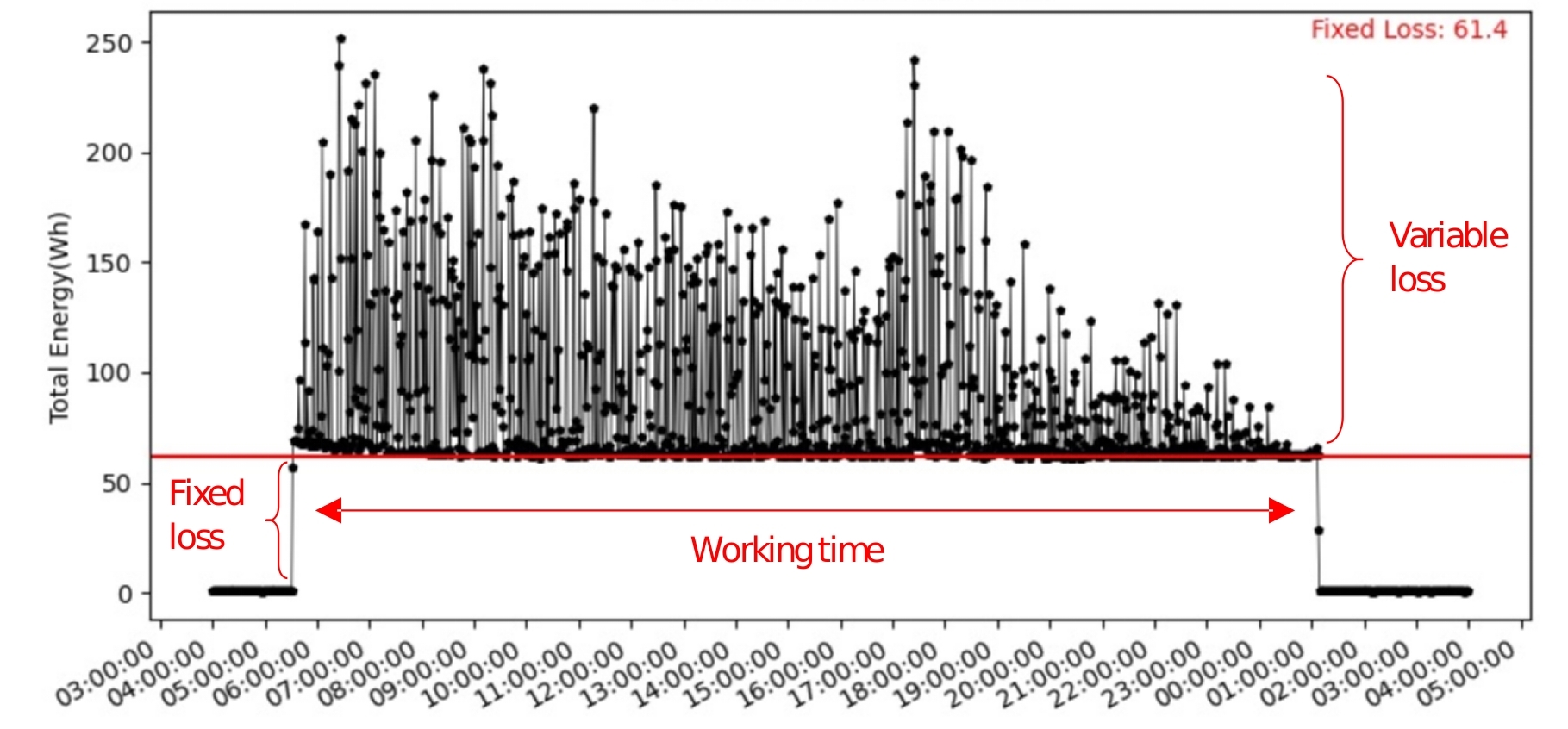}
\caption{\label{fig:energy_consump_plot}Energy consumption plot of Escalator 19 on a single day.}
\end{figure}

Under the same conditions, the fixed loss energy on upward escalators is always higher than the downward ones. Meanwhile, the fixed loss energy is higher for escalators with a higher rise, as they need more power to run. To estimate the workload we use the following equation, derived from \cite{al2011modelling}:


\begin{equation}
P_{d} = \frac{E_{V}*3600}{g r_{e} m_{p} k_{wf}},
\end{equation}

\noindent where $P_d$ is the daily estimated passenger count, \(g=9.81m/s^{2}\) is gravitational acceleration, \(r_{e}\) is the vertical rise of an escalator, \(m_{p}=75kg\) is the average mass of a passenger, \(k_{wf}\) is the walking factor, and 1Wh=3600J. We use \(k_{wf}=0.75\) for downward escalators and \(k_{wf}=0.85\) for upward escalators. Figure \ref{fig:energy_consump_plot} illustrates the working time, fixed and variable energy loss on daily energy consumption. 

\begin{figure}
     \centering
     \begin{subfigure}[b]{0.48\textwidth}
         \centering
         \includegraphics[width=\textwidth]{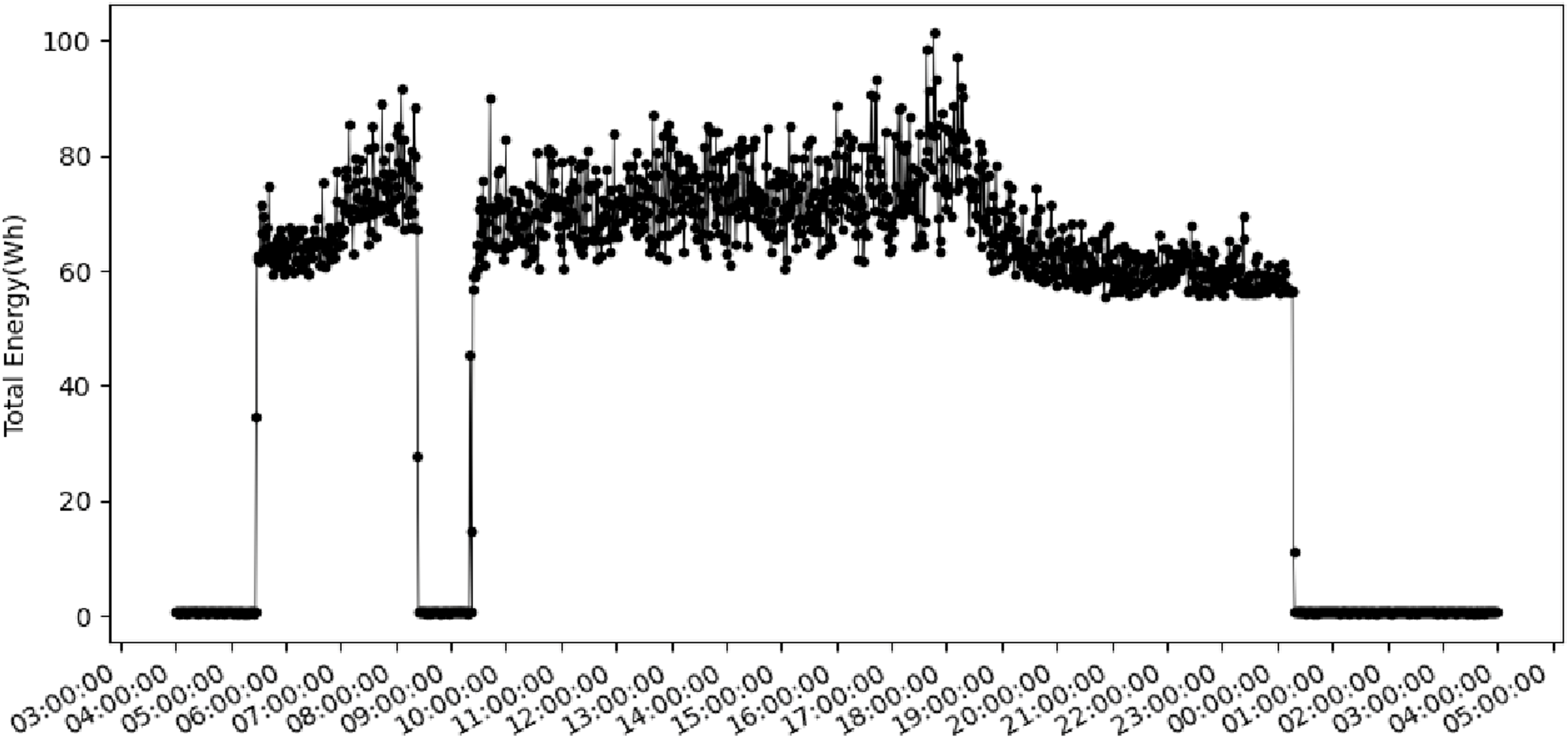}
         \caption{corrective maintenance}
         \label{fig:energy_consump_events_main}
     \end{subfigure}
     \hfill
     \begin{subfigure}[b]{0.48\textwidth}
         \centering
         \includegraphics[width=\textwidth]{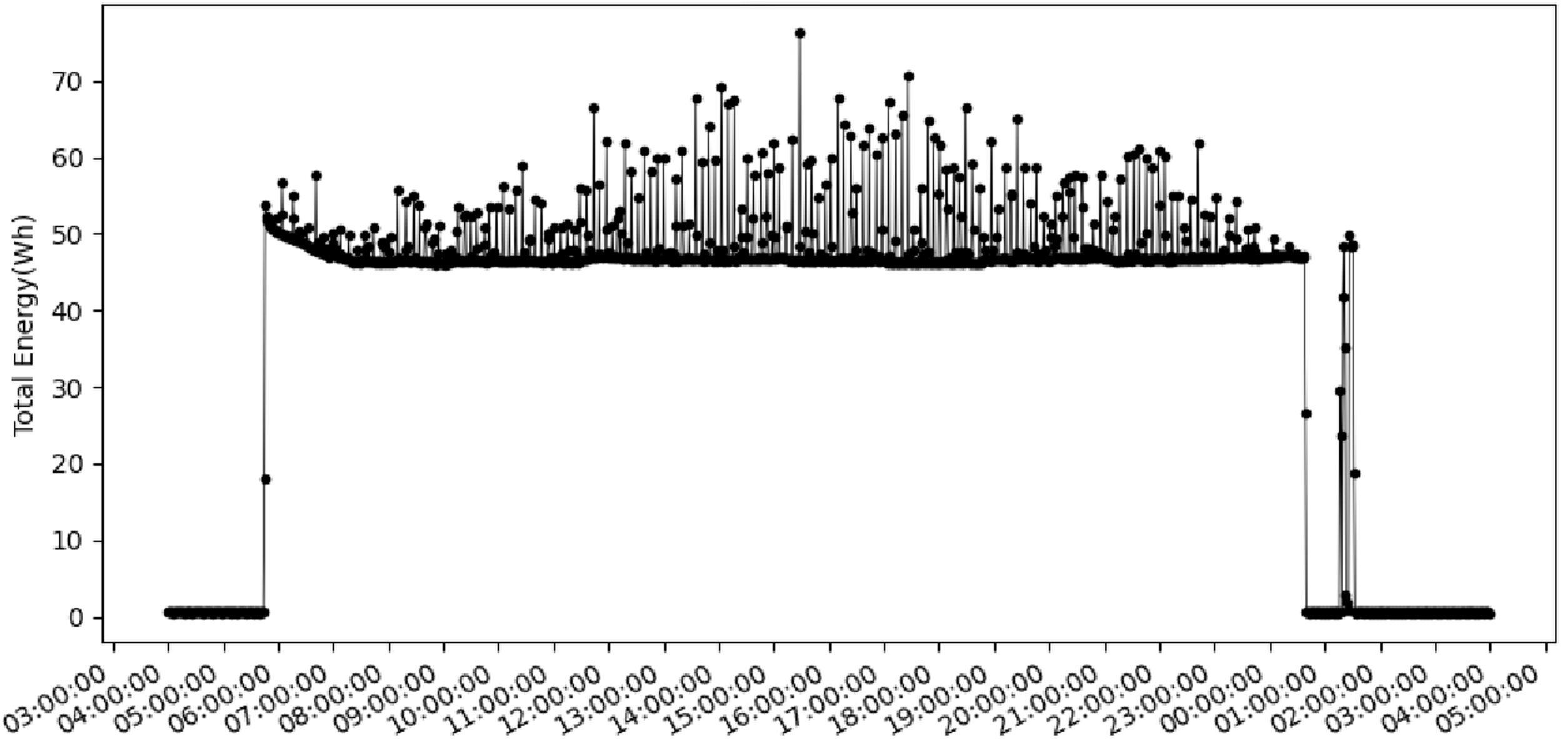}
         \caption{preventive maintenance}
         \label{fig:energy_consump_events_preven}
     \end{subfigure}
        \caption{Energy consumption with maintenance events.}
        \label{fig:energy_consump_events}
\end{figure}

During a regular operating period, zero or tiny (less than 5 Wh) energy consumption always indicates an unexpected escalator shutdown, which is highly related to a fault. And corrective maintenance will be done after the fault. Conversely, energy consumption during the regular non-service time usually relates to scheduled preventive maintenance, as the escalator should be turned on after a preventive maintenance event for a check. Thus, the algorithms can identify corrective and preventive maintenance by the daily energy consumption. Figure \ref{fig:energy_consump_events} shows examples of corrective and preventive maintenance events, respectively. Both processed and raw data are stored in a database.

\subsection{Dashboard and monitoring}

\begin{figure}[h!t]
\centering
\includegraphics[width=0.9\textwidth]{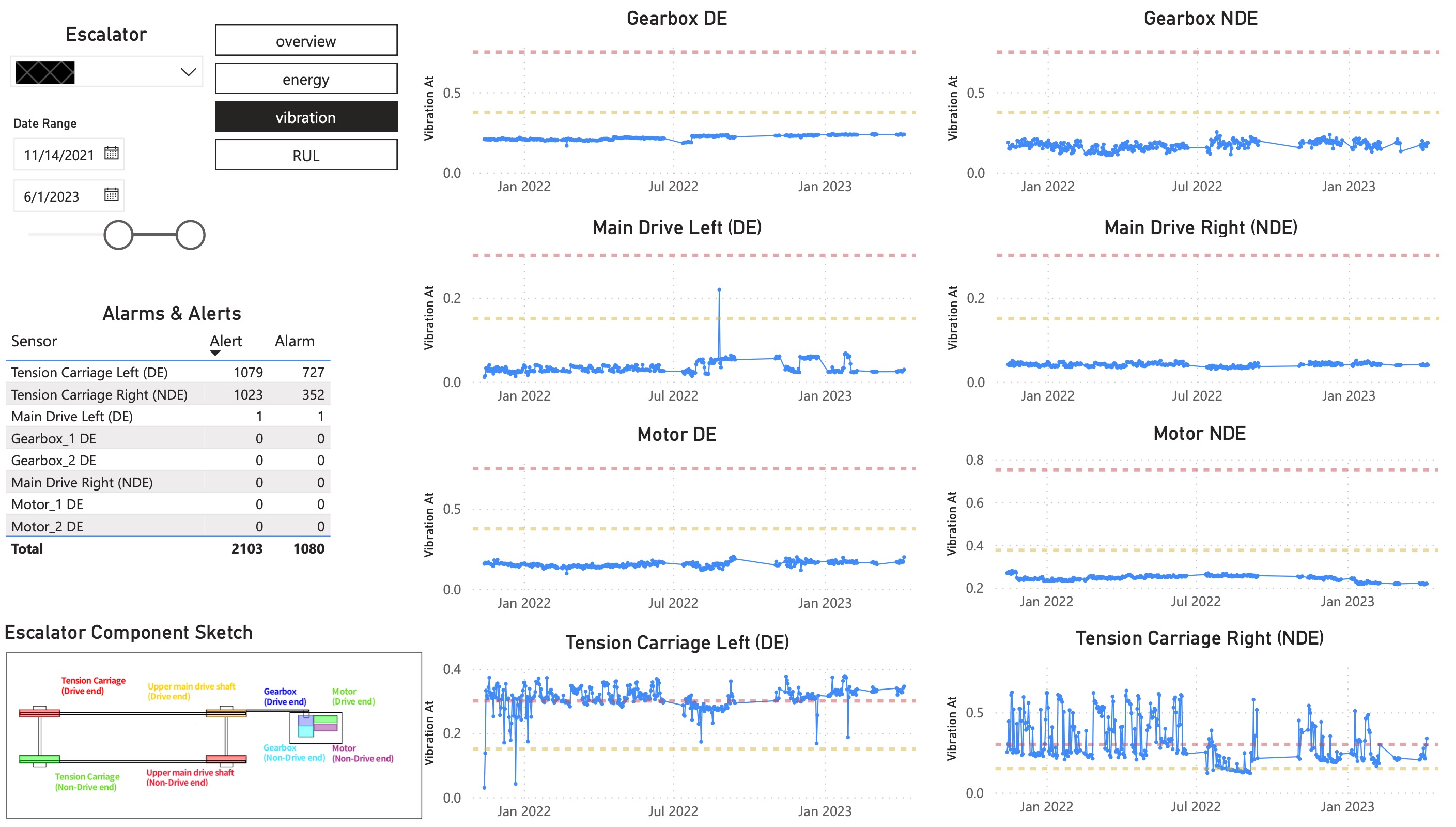}
\caption{\label{fig:vibration_dashboard}Vibration sheet of the dashboard.}
\end{figure}

Various kinds of data are used for monitoring and displayed on the dashboard, i.e., daily fixed loss, daily passenger count, daily working hours, $A_{t}$ alarm and alert count. The dashboard mainly consists of four sheets, i.e., overview, energy, vibration and RUL sheets. The overview sheet contains the basic information and alarm and alert count during the selected period of specific escalators. The energy-related information is shown in different graphs, working time, passenger load and fixed energy loss on the energy sheet. $A_{t}$ values, alarms and alerts of each sensor of different escalator sensors are visualized on the vibration sheet. Finally, the quarterly LHI and RUL predictions are visualized on the RUL sheet. The vibration sheet is illustrated in Figure \ref{fig:vibration_dashboard}.

\section{Health index}

The lifetime health index (LHI) \(y\) indicates the health status of the escalator in a specific time period. In this project, the LHI of each escalator is computed quarterly. Without loss of generality, we assume the LHI \(y\) is computed in a quarter \(q\) from date \(d_{1}\) to \(d_{2}\), where all the data are available. It is assumed that the elder escalators have the higher LHI, as they are usually in worse operating status. 

To study the relationship between the LHI \(y\) and the escalator age \(t\), we first compute the LHI \(y\) as the weighted sum of five standardised variables: working time \(T\), passenger load \(P\), fixed loss residuals \(L\), \(A_{t}\) exceedance curve area \(N\) and fault count \(C\); and the larger \(y\), the worse the health status of an escalator, as shown in the following formula.

\begin{equation}
    y=\omega_{1}*T+\omega_{2}*P+\omega_{3}*L+\omega_{4}*N+\omega_{5}*C,
\end{equation}

\noindent where \(\omega_{i}\) are the weights, and \(\omega_{1}=0.2\), \(\omega_{2}=0.2\), \(\omega_{3}=0.2\), \(\omega_{4}=0.3\), and \(\omega_{5}=0.1\). These five factors can be divided into long-term and short-term. The former describes the accumulated performance, like working time and passenger load, which indicates the degradation of the whole escalator mechanical system. The latter shows the escalator performance in a period like fixed loss residuals, fault count and \(A_{t}\) alarm count, which is influenced by human factors like maintenance events.

The five variables are the processed data from the pre-processed steps discussed in Section \ref{section:data_pipeline}. The processed data are then normalised by the min-max scalar. The exact minimum and maximum values of each variable are listed in Table \ref{table:parameters_lhi}, along with their weights. 

\begin{table}[h!]
\centering
\begin{tabular}{c c c c c c} 
 \hline \hline
\multirow{2}{*}{Variables} & Working & Passenger & Fixed loss & Exceedance & Fault \\ 
& time \(T\) & load \(P\) & residual \(L\) & curve area \(N\) & count \(C\) \\ [0.5ex] 
\hline
Weights & 0.2      & 0.2       & 0.2   & 0.3  & 0.1 \\ [0.2ex]
Minimum & 0        & 0         & 0     & 0    & 0   \\ [0.2ex]
Maximum & 15330000 & 332150000 & 19.61 & 0.15 & 33  \\ [0.5ex]
\hline \hline
\end{tabular}
\caption{Parameters to compute the normalized variables of the LHI value.}
\label{table:parameters_lhi}
\end{table}

\begin{figure}[h!t]
\centering
\includegraphics[width=0.7\textwidth]{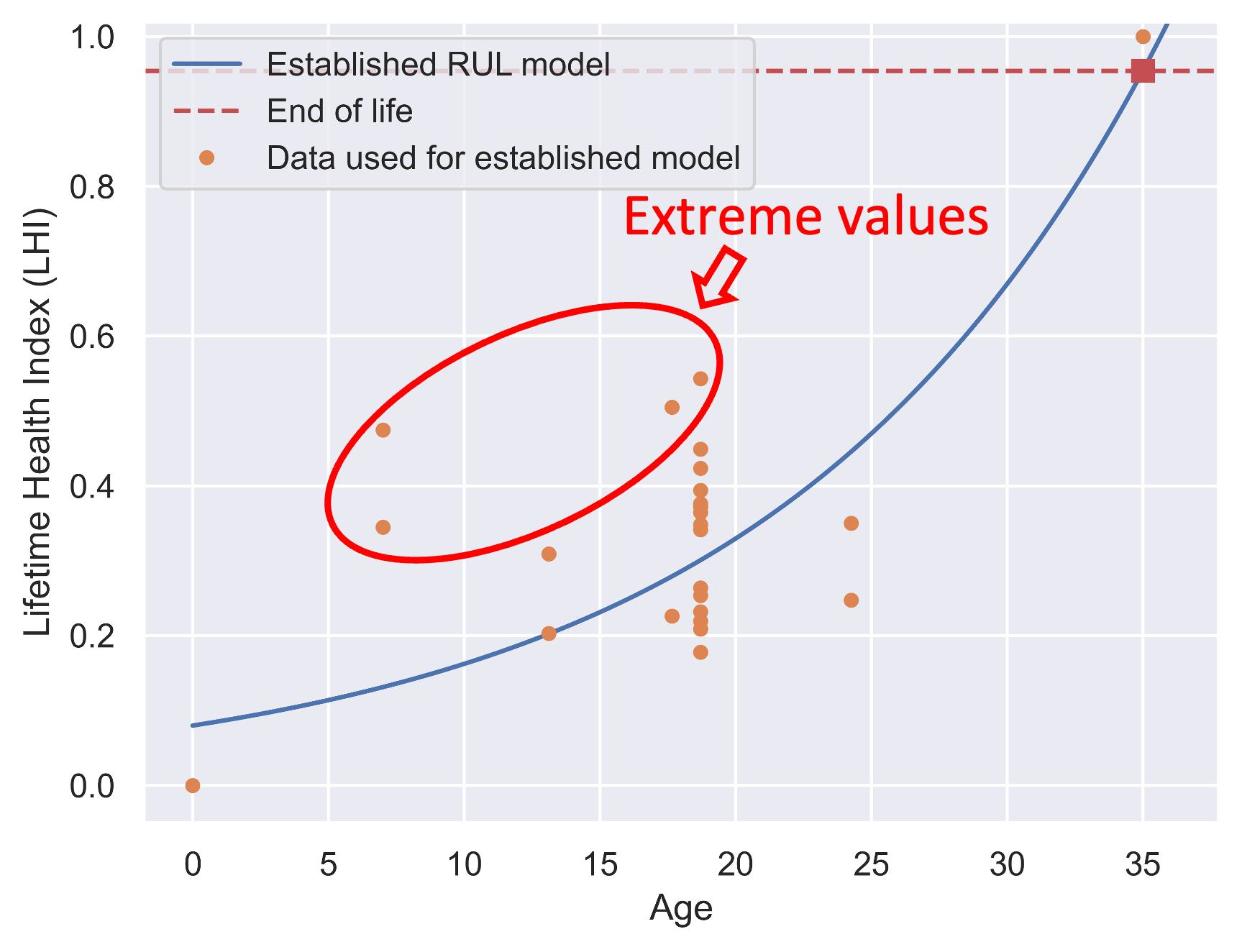}
\caption{\label{fig:lhi_plot}Scatter plot of the LHI values and fitted line for the LHI model.}
\end{figure}

The normalized data are utilized to compute the LHI of escalators. The model $F$, indicating the relationship between the escalator LHI $y$ and age $t$, is built by all escalators except four with extreme LHI values, (see Figure 20). The model is fit as,

\begin{equation}
    y=F(t)=a \cdot exp(b \cdot t), \label{eq:lhi_equa}
\end{equation}

\noindent where \(a=0.0928\) is the starting value, \(exp(b)=exp(0.0665)=1.069\) is the LHI growth rate per year. The LHI model \(F\) is obtained by fitting an exponential curve from the LHI values of the escalators in the project so that it serves as the reference model. If we apply model \(F\) to a specific escalator, it should be shifted horizontally to pass through the LHI value of the escalator. Figure \ref{fig:lhi_plot} illustrates the fitted curve of the model \(F\) with all escalators' LHI values. 

\section{RUL model}

Based on the LHI model (i.e., Eq. \eqref{eq:lhi_equa}), the RUL of an escalator could be derived. To achieve this, the estimated age \(\hat{T}\) of the escalator is first computed by, 

\begin{equation}
    \hat{T}=F^{-1}(y), \label{eq:rul_equa}
\end{equation}

\noindent where \(y\) is the LHI value of the escalator, and \(F^{-1}\) is the inverse function of the LHI model \(F\). Unlike the actual age \(T\) that only counts the service time of the escalator, the estimated age \(\hat{T}\) considers the health condition of the escalator and is regarded as a more robust indicator to compute RUL.

Now for escalator \(e\), denote the actual age by \(T_{e}\) and the estimated age by \(\hat{T}_{e}\) then the \(RUL_{e}\) can be computed by,

\begin{equation}
    RUL_{e} = T_{end} - \hat{T}_{e}, \label{eq:rul_equa_esca}
\end{equation}

\noindent where \(T_{end}\) is defined as 35 from the industrial practice.

\begin{figure}[!ht]
\centering
\includegraphics[width=0.7\textwidth]{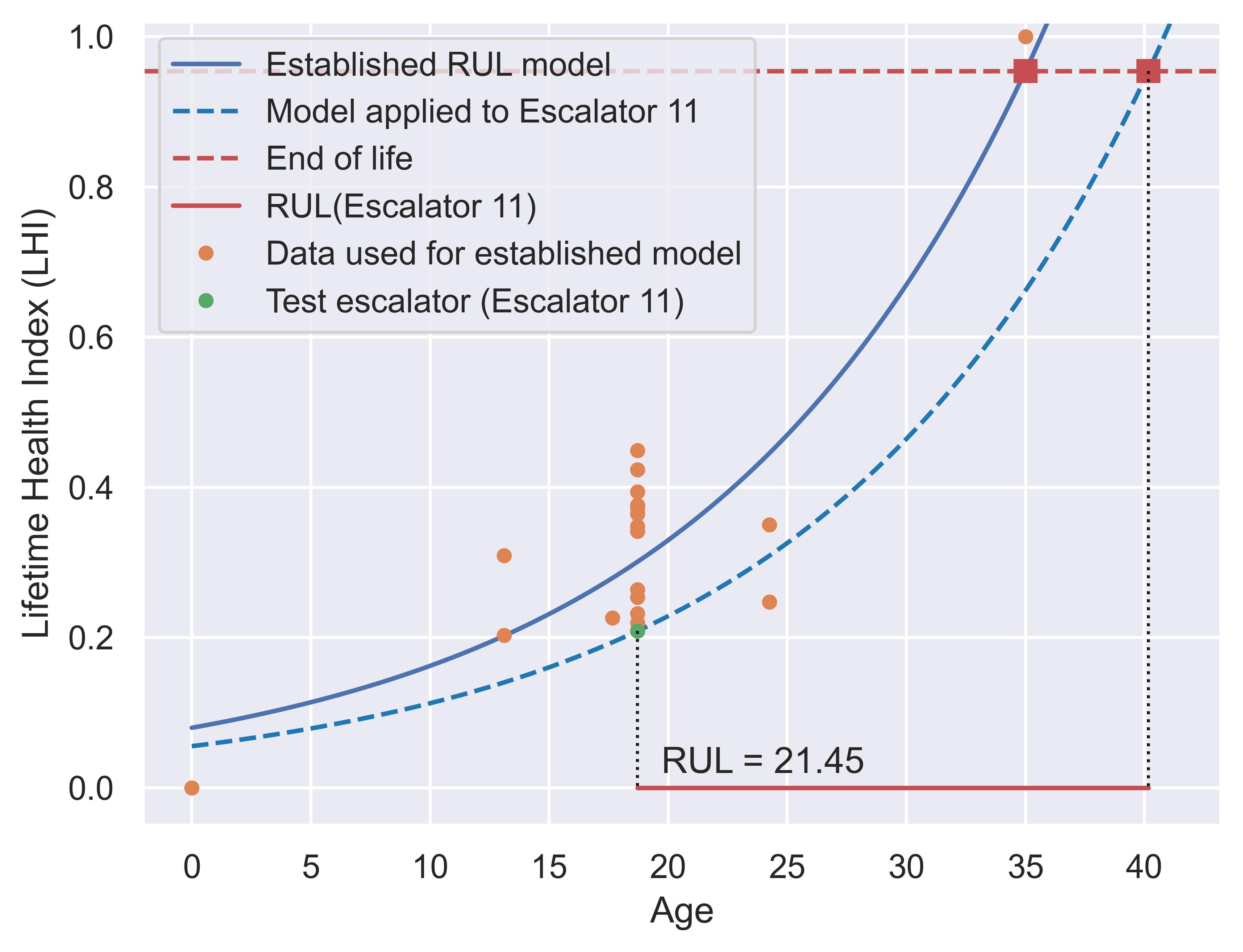}
\caption{\label{fig:rul_plot}RUL plots for Escalator 11.}
\end{figure}

Figure \ref{fig:rul_plot} illustrates the RUL of Escalator 11. Given the LHI value \(y_{11}\) of Escalator 11, an 18.7-year-old escalator. We first horizontally move the reference model \(F\) (in a solid blue line) passing through its LHI value (green dot). Moreover, this curve is cut down when it reaches the LHI threshold value \(y_{end}\), and the corresponding age \(\hat{T}_{end}\) will be the ending life end Escalator 11. The difference between the estimated ending service time, \(\hat{T}_{end}\), and the actual service time, \(T\), is the remaining useful life of Escalator 11, i.e., \(RUL_{11}=40.15-18.7=21.45\) Years.

\begin{table}[!ht]
\centering
\resizebox{\columnwidth}{!}{\begin{tabular}{c c c c c c c c c c c c} 
 \hline \hline
escalator & \multirow{2}{*}{Year} & Quar- & Actual & Year & \multirow{2}{*}{RUL} & \multirow{2}{*}{LHI} & Working & Passenger & \(A_{t}\) & Fixed Loss & Fault \\ 
id & & ter & Age & Till 35 & & & Hours & Load & Areas & Residual & Counts \\ [0.5ex] 
\hline
0  & 2021 & 4 & 6.34  & 28.66 & 14.82 & 0.36 & 0.18 & 0.16 & 0.90 & 0.22 & 0.24 \\ 
1  & 2021 & 4 & 6.34  & 28.66 & 17.68 & 0.29 & 0.18 & 0.15 & 0.74 & 0.08 & 0.03 \\  
2  & 2021 & 4 & 23.59 & 11.41 & 16.65 & 0.31 & 0.66 & 0.31 & 0.36 & 0.00 & 0.12 \\  
3  & 2021 & 4 & 23.59 & 11.41 & 21.48 & 0.23 & 0.67 & 0.28 & 0.10 & 0.00 & 0.06 \\ 
4  & 2021 & 4 & 18.05 & 16.95 & 21.36 & 0.23 & 0.48 & 0.14 & 0.22 & 0.15 & 0.09 \\  
5  & 2021 & 4 & 18.05 & 16.95 & 18.68 & 0.27 & 0.33 & 0.16 & 0.19 & 0.55 & 0.12 \\  
6  & 2021 & 4 & 12.44 & 22.56 & 18.34 & 0.28 & 0.37 & 0.19 & 0.55 & 0.00 & 0.06 \\ 
7  & 2021 & 4 & 12.44 & 22.56 & 25.51 & 0.17 & 0.37 & 0.16 & 0.23 & 0.00 & 0.00 \\  
8  & 2021 & 4 & 18.05 & 16.95 & 23.11 & 0.20 & 0.38 & 0.20 & 0.20 & 0.14 & 0.00 \\  
9  & 2021 & 4 & 18.05 & 16.95 & 15.14 & 0.35 & 0.49 & 0.28 & 0.37 & 0.40 & 0.03 \\  
10 & 2021 & 4 & 18.05 & 16.95 & 27.53 & 0.15 & 0.06 & 0.03 & 0.20 & 0.36 & 0.03 \\ 
11 & 2021 & 4 & 18.05 & 16.95 & 24.27 & 0.19 & 0.49 & 0.08 & 0.19 & 0.00 & 0.18 \\ 
12 & 2021 & 4 & 18.05 & 16.95 & 14.60 & 0.36 & 0.50 & 0.41 & 0.10 & 0.64 & 0.21 \\ 
13 & 2021 & 4 & 18.05 & 16.95 & 19.27 & 0.26 & 0.50 & 0.35 & 0.09 & 0.11 & 0.06 \\ 
14 & 2021 & 4 & 18.05 & 16.95 & 14.25 & 0.37 & 0.51 & 0.50 & 0.24 & 0.45 & 0.06 \\ 
15 & 2021 & 4 & 18.05 & 16.95 & 16.85 & 0.31 & 0.51 & 0.37 & 0.39 & 0.00 & 0.18 \\ 
16 & 2021 & 4 & 17.00 & 18.00 & 25.15 & 0.18 & 0.37 & 0.13 & 0.18 & 0.06 & 0.12 \\ 
17 & 2021 & 4 & 17.00 & 18.00 & 16.07 & 0.33 & 0.37 & 0.11 & 0.39 & 0.54 & 0.06 \\ 
18 & 2021 & 4 & 18.05 & 16.95 & 17.68 & 0.29 & 0.48 & 0.40 & 0.27 & 0.10 & 0.18 \\ 
19 & 2021 & 4 & 18.05 & 16.95 & 11.25 & 0.45 & 0.48 & 0.34 & 0.43 & 0.72 & 0.15 \\ 
20 & 2021 & 4 & 18.05 & 16.95 & 22.66 & 0.21 & 0.46 & 0.22 & 0.20 & 0.07 & 0.00 \\ 
21 & 2021 & 4 & 18.05 & 16.95 & 15.36 & 0.34 & 0.50 & 0.26 & 0.42 & 0.28 & 0.09 \\ 
22 & 2021 & 4 & 18.05 & 16.95 & 15.16 & 0.35 & 0.50 & 0.38 & 0.47 & 0.14 & 0.03 \\ 
23 & 2021 & 4 & 18.05 & 16.95 & 18.84 & 0.27 & 0.50 & 0.13 & 0.20 & 0.40 & 0.03 \\ [0.5ex] 
\hline \hline
\end{tabular}}
\caption{Remaining Useful Life (RUL) for 24 escalators.}
\label{table:rul_all_esca}
\end{table}

The RUL modelling results are provided in detail in Table \ref{table:rul_all_esca}, which shows the RUL of all escalators in 2022Q1 and the LHI with five variables.

\section{Overview and future work}

We have developed a Health Condition Analytics System for escalators that can be used for condition monitoring and remaining useful life predictions. These predictions inform refurbishment planning decisions and enable the reliable extension of the useful life of well-functioning escalators beyond the recommended lifetime thereby reducing waste. The model uses real-time measurement data from various vibration and energy sensors. Thereby incorporating the current operating condition (vibration) as well as the workload (based on energy consumption) into the model. 

The model may be expanded to include more accurate data on faults when that is available. In addition, the verification of the model predictions will have to be done in time as escalators become older and get closer to their end of useful life. Though the system has been designed for escalators, it is applicable to other engineering systems where energy consumption can be used to approximate workload and vibration is an important indicator of wear and tear. 
\newpage
\section*{Acknowledgement}

The work in this paper was supported by a grant from the Innovation and Technology Fund (ref: PRP-008-20FX) and MTR Corporation Limited. This work is related to Hong Kong Short-Term Patent Application No. 32023068734.7.


\end{document}